\documentclass[A4paper,pre,twocolumn,showpacs,aps,floats,floatfix,superscriptaddress]{revtex4}

\usepackage{graphicx}
\usepackage{graphics}
\usepackage{subfigure}
\usepackage{epsfig}
\usepackage{amsmath,amssymb,bm}

\usepackage[colorlinks,linkcolor=blue,citecolor=blue]{hyperref}
\usepackage{dcolumn}
\usepackage{rotating}
\usepackage{multirow}
\usepackage{color}
\usepackage{float}
\usepackage{setspace}
\usepackage{natbib}

\begin{document}

\title{Community core detection in transportation networks}
\author{Vincenzo De Leo} \affiliation{Linkalab,
  Complex Systems Computational Laboratory, Cagliari, Italy} 
\author{Giovanni Santoboni} \affiliation{AECOM, 2201 Wilson Boulevard, 
  Suite 800, Arlington VA, 22201, USA} 
\author{Federica Cerina} \affiliation{Linkalab, Complex
  Systems Computational Laboratory, Cagliari, Italy} \affiliation{Department of Physics,
  University of Cagliari, Italy}  \affiliation{IMT Institute 
  for Advanced Studies Lucca, Piazza S. Ponziano 6, 55100 Lucca, Italy}
\author{Mario Mureddu} \affiliation{Linkalab, Complex
  Systems Computational Laboratory, Cagliari, Italy} \affiliation{Department of Physics,
  University of Cagliari, Italy}  \affiliation{IMT Institute 
  for Advanced Studies Lucca, Piazza S. Ponziano 6, 55100 Lucca, Italy}
\author{Luca Secchi} \affiliation{Linkalab, Complex
  Systems Computational Laboratory, Cagliari, Italy} 
\author{Alessandro Chessa} \affiliation{Linkalab, Complex
  Systems Computational Laboratory, Cagliari, Italy}  \affiliation{IMT Institute 
  for Advanced Studies Lucca, Piazza S. Ponziano 6, 55100 Lucca, Italy} \affiliation{Institute 
  for Complex Systems, CNR UOS Department of Physics, University of 
  Rome "La Sapienza", Rome, Italy}

\begin{abstract}
This work analyses methods for the identification and the stability under perturbation of a territorial community structure with specific reference to transportation networks. 
We considered networks of commuters for a city and an insular region. In both cases, we have studied the distribution of commuters' trips ({\it i.e.}, home-to-work trips and viceversa).
The identification and stability of the communities' cores are linked to the land-use distribution within the zone system, and therefore their proper definition may be useful to transport planners.
\end{abstract}


\maketitle 

\section{Introduction}

Many Complex Systems can be modelled as networks, in which vertices are the entities of interest in the system under investigation and edges are the relations between couple of vertices/entities. For example in the World Wide Web the vertices are the web pages and the edges are the hyperlinks (in this case the network is directed and we have arcs instead of simple edges).
Intuitively, not all vertices and edges have equal roles within a large-scale network; some vertices may be of some importance for the distribution of traffic in the network, and the edges that carry most of the traffic do so because they connect "groups" of vertices that are particularly important within the network.
The scope of this paper is to understand the nature of these "groups", their "community structure" or "clustering", and find ways to determine the importance of vertices inside each community, revealing its inner hierarchy.
The community structure of a network is a topic that has been comprehensively treated in \cite{Fortunato:2010}. 

The first problem of graph clustering is one of definition. 
Although the concept is intuitive, it is not defined in a rigorous way, as there is no definition of community boundary, or a unique way of determining whether a particular edge is part of a community and not of another. 
Therefore, as pointed out in \cite{Fortunato:2010}, communities are algorithmically defined, {\it i.e.}, they are the final product of the algorithm, without a precise {\it a priori} definition.

The field of transportation is a natural choice for the definition of a community structure, though the field itself has some inherent limitations. 
On a practical matter, the measurement of important traffic variables is lengthy and expensive. 
For once, different methods to count traffic volumes return different answers, especially in the identification of commercial vehicles \cite{fhwa}. 
Additionally, the development of a regionwide origin-destination (OD) matrix at the zone level is a long and costly procedure; in particular the matrix of the metropolitan area used in this study has been derived after a year-long survey process, and the final OD matrix is assembled by weighting a matrix of survey responses according to the population of the areas where the partecipants live.
A second calibration stage is generally done to test whether the OD matrix obtained assignes traffic compatibly with the traffic on the major highways of the study area; as a result of this process, the trip distribution and assignment may work well {\em globally}, but larger discrepancies may persist {\em locally}.
Finally, during the time occurred to carry out this process, conditions on the ground may have already changed, since the land-use of an area is constantly changing, therefore creating discrepancies in the final OD matrix. 

Notwithstanding these inherent difficulties, the identification of communities within a metropolitan area network still holds great importance. 
First, the formation of communities in a network is a byproduct of land-use development.
Land-use development occurs for a number of reasons (service maximization, profit, etc), and the location for development is chosen according to the optimization in terms of different variables, like price of land, proximity to transit, regulation, that are however variables related to each zone/vertex of the system. 
For example, demand for transport between two vertices may lead to the opening of a new edge ({\it e.g.}, a new bus route, a new road), which in turn may lead to more demand for transport (in the form of "induced demand", \cite{mishan,pashigian}). 
The community structure is not solely a function of the attributes of each zone/vertex, but also of the network arrangement, hence it forms a more comprehensive measure of the importance of a group of zones as a subsection of the zone system.

It is important to know which vertices are the most relevant from the point of view of the internal stability of a community and the overall partition structure.
We will see in the next section that this idea is at the cornerstone of the community stability.
In other fields the problem has been studied in terms of network breakdown, which has found applications in the accessibility of a transportation network for flood damage.
Knowledge of community structure can serve planners in the situation of natural disasters to predict the onset of network breakdown, as studied in \cite{flood}.
In other fields, it has been applied to the identification of crucial edges in a web network under cybernetic attack \cite{Albert_et:2000,Sole_et:2008,Schneider_et:2011}.

\section{Materials and Methods}

\subsection{Community detection and modularity}

There are now many community detection methods \cite{Fortunato:2010} 
and the most popular is the modularity optimization introduced by Newman
and Girvan \cite{Newman:2004}. 
This method has various drawbacks, the most important of which is the existence of a resolution limit \cite{Fortunato:2007} which prevent it to detect smaller modules, but has also the advantage of being easy to implement.
The modularity function that needs to be optimized is defined as \cite{Newman:2006}:
\begin{equation}
Q=\frac{1}{2m}\sum_{ij}\left(A_{ij}-P_{ij}\right)\delta(C_i,C_j)
\label{modularity}
\end{equation}
where the sum is over all the node pairs, $A$ is the adjacency matrix, $m$ is the 
total number of edges and $P_{ij}$ is the expected number of edges between the 
vertices $i$ and $j$ for a given null model. 
The function will result in a null contribution for couples of vertices not belonging to the same community ($C_{i} \neq C_{j}$). 
For an unweighted network, the choice $P_{ij} = k_{i}k_{j}/2m$ equates to taking as a null model a random network with the same degree sequence as the original network.

To optimize the modularity we used the Louvain algorithm \cite{Blondel:2008} based on two steps that are repeated iteratively until a global maximum is reached.
In the first step a network partition in which the amount of communities is equal to the amount of nodes is created. 
Then, the algorithm iterates over all nodes and computes for each node the modularity gain within the communities of its neighbors; a node movement is maintained if it leads to a positive variation in modularity.
The iteration is repeated until a local maximum is reached, that is until there is not any other move that lead to an increase in modularity. 

In the second step the algorithm creates a new network whose nodes are the communities detected at the end of the first step; the weight of 
the new links between these new nodes is the total weight of the links between the old nodes that belong to the communities they come from.
Typically the amount of nodes decrease drastically at this step and this ensures a fast convergence of the algorithm for large networks. 

The main problems of all algorithms for community detection is the fact that the community definition does not provide any information about the importance of a node inside its own community. 
Nodes of a community do not have all the same importance for the community stability: the removal of a node in the "core" of a network affects the partition much more than the
deletion of a node that stays on the edge of the community ({\it i.e.} a node connected in the same
way with nodes internal and external to its community).
The purpose of the following section is to develop a novel way for detecting cores inside communities by using the properties the of modularity function.

\subsection{$dQ$ analysis for cores detection in a partition}
By definition, if the modularity associated to a network has been optimized,
every perturbation in the partition leads to a negative variation in the modularity
($dQ$). 

\begin{figure}[h!]
\centering
  \includegraphics*[width=8.6cm]{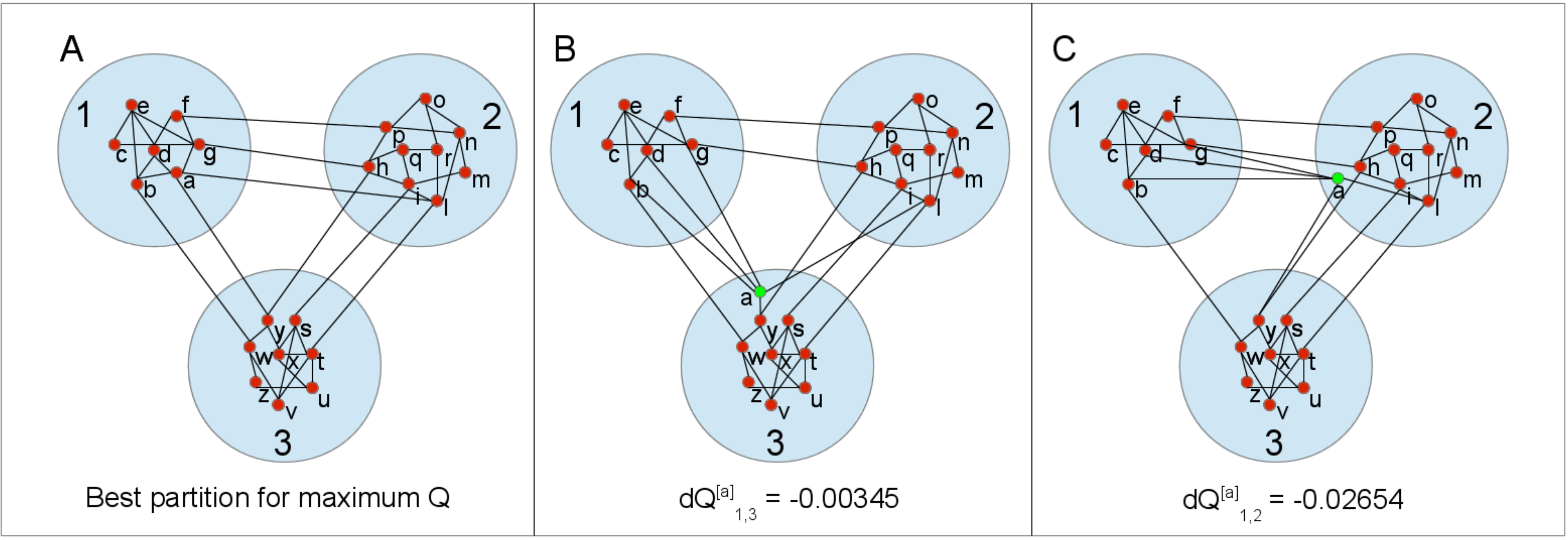}
  \caption{This picture (A) describe the situation in which the modularity of a network has been maximized. Starting from this state 
    it is possible to determine the list of negative dQ values associated to each node assuming to move it in every other community. As a matter of fact,
    if a node (in this picture we consider as example the node 'a') would change its belonging to the community in which has been placed during the modularity
    optimization, the modularity of the network would obviously decrease, as shown in (B) and (C). This negative variation is related to the fact that, for each 
    change in the partition, like the ones depicted in (B) and (C), the total number of links internal to the communities is always smaller with respect 
    to the one associated to (A).}
  \label{fig:example}
\end{figure}

If we move a node from its community we have $M-1$ possible choices (with M the number of communities) as possible targets for the new host community of this node.
We decided to define the $dQ$ associated to each node as the smallest variation in absolute value (or the closest to $0$ since $dQ$ is always a negative number) for all the possible choices and this is in our view a measure of how that node is internal in its community.

\begin{figure}[h!]
\centering
  \includegraphics[width=8.6cm]{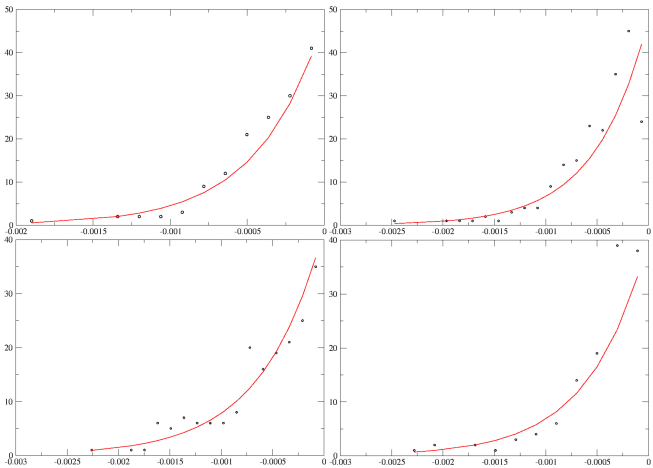}
  \caption{$dQ$ frequency plots relative to 4 communities detected for the city of Atlanta, GA.
    The correlation coefficients of the exponential fits are (from top right to bottom left, respectively) 0.956, 0.946, 0.937 and 0.933. 
    In general, these distributions are the tipical $dQ$ frequency distribution inside a community (provided there are enough nodes to perform an exponential fit).}
  \label{fig:dQ_distribution}
\end{figure}
Fig. \ref{fig:dQ_distribution} shows the typical $dQ$ frequency distribution of nodes inside a community; the data points were fitted using a decaying exponential form $\exp(-x/\ell)$ with typical length $\ell$. 
The typical lenght $\ell$ and defines a starting point to discriminate the core nodes. For practical purposes, the threshold value $d_{thr}=2\ell$ is an appropriate boundary value to differentiate between core nodes (the ones below the threshold) and the border nodes (the peripheral nodes). With this choice we found that, for what it concerns the networks described in this work, the percentage of core nodes is, for every community of every network, always equal to the 8\% of the total amount of nodes in that particular community.

\begin{figure}[h!]
\centering
  \includegraphics[width=9.0cm]{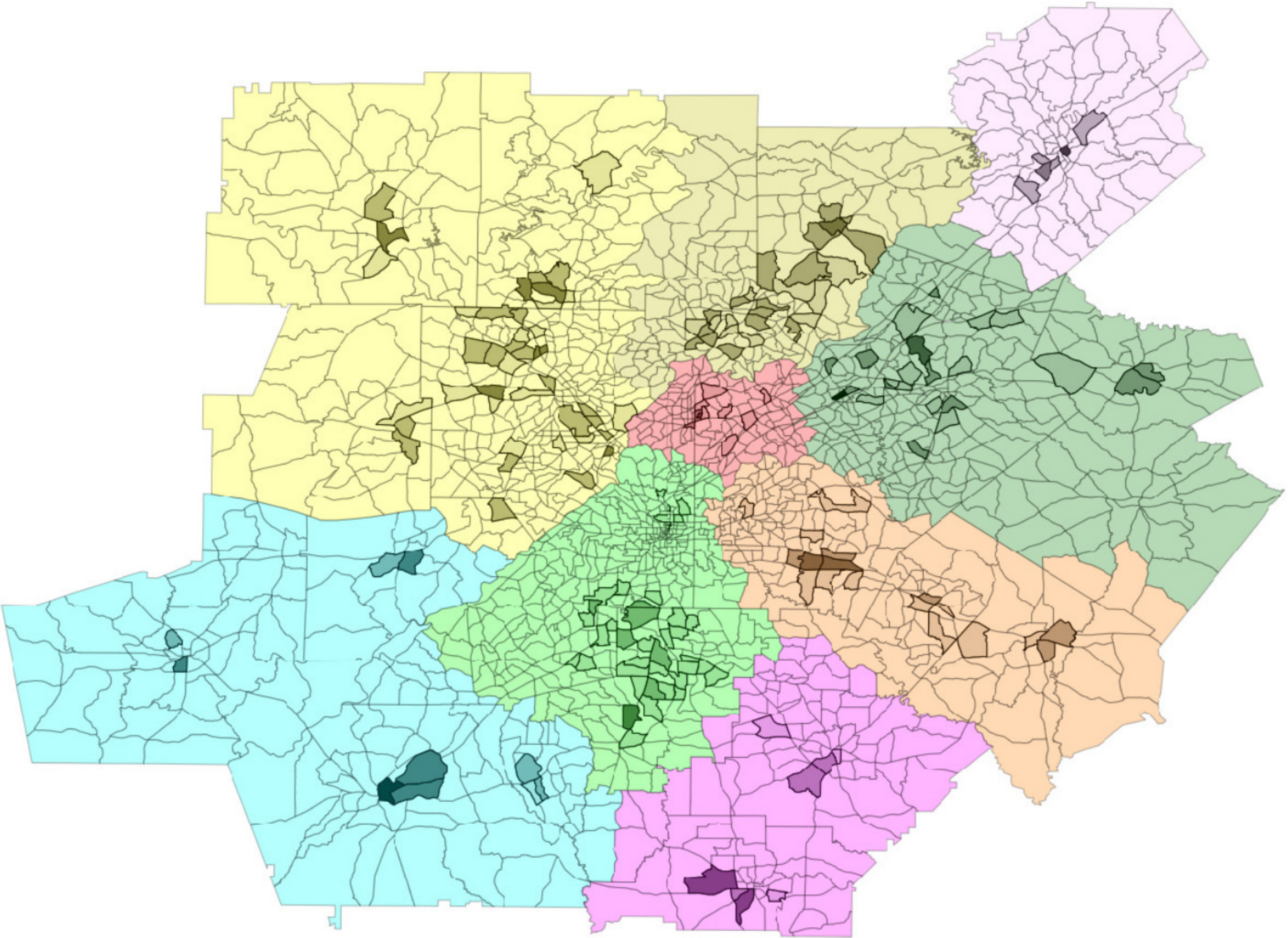}
  \caption{Cores detected for the city of Atlanta, GA, using a threshold equal to double the typical length of the exponential distribution of the dQ frequencies.}
  \label{fig:core}
\end{figure}

Fig. \ref{fig:core} shows the cores detected for the city of Atlanta, GA, using the method described above. The nodes of the network correspond to the TAZ (Traffic Analysis Zone) of the city and the links' weight have been computed summing, for each couple of TAZ, the corresponding traffic flow in both directions, as described in more detail later.

\section{Datasets}
\label{sec:Datasets}
This paper analyses methods for the identification and the stability of a community structure using two networks from the field of transportation. The first network is a regionwide network of commuting trips in the insular region of Sardinia, in Italy, the Sardinian Inter-municipal Commuting Network (SMCN).
The second network is a network of daily commuting trips in the metropolitan area of Atlanta, USA, part of the Atlanta Regional Commission (ARC) model. 
In both cases, we have studied the distribution of commuting trips, {\it i.e.}, home-to-work trips and viceversa. 
The choice was determined by the fact that trips of these types are clearly defined to planners, because their correlation to the land-use is well understood, necessarly tied to the population of the origin zone and the employment of the destination zone.

The trip distribution in both zone systems have in common the fact of being derived through interpolation of a survey. 
In short, a questionnaire was used to obtain a set of origin-destination movements per purpose of travel, which was then expanded according to the population of the various zones. 
The trips are differentiated by direction of travel, hence trips $i\to j$ and $j\to i$ are different.

The main difference in the two zone systems is due to the extension of the zones. While in the SMCN the zones have the dimension of a municipality, in the ARC the size of the zones changes according to the structure of the road network. In the downtown areas of Atlanta these zones may even have the size of a traffic block. 

\subsection{Sardinian Inter-municipal Commuting Network}
\begin{figure}[ht!]
\centering
\includegraphics*[width=0.18\textwidth]{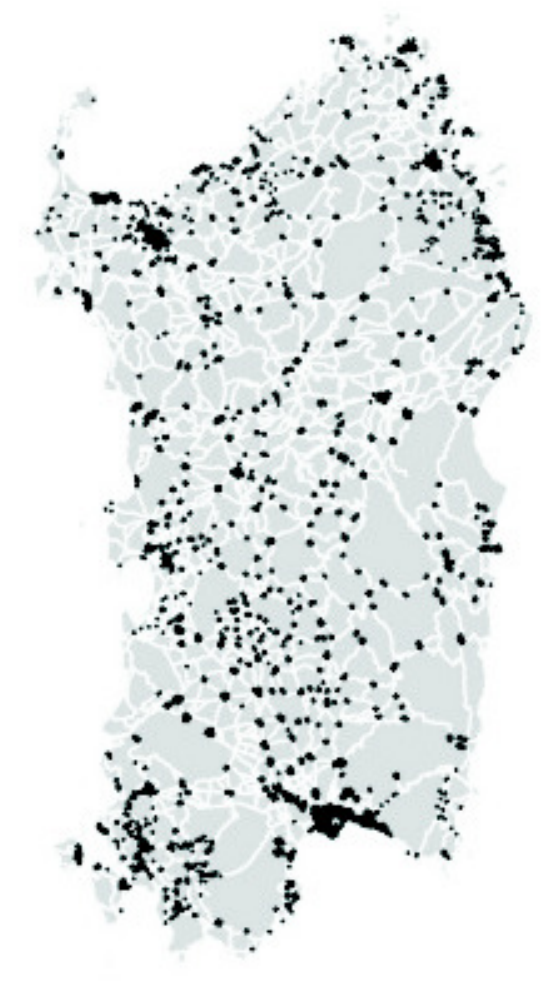}\includegraphics*[width=0.28\textwidth]{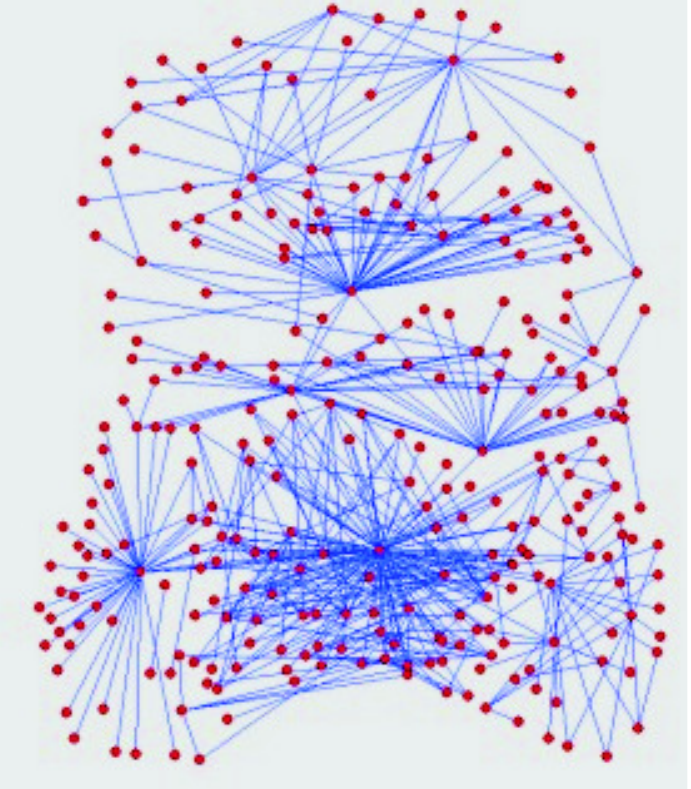}
  \caption{Geographical versus topologic representation of the the Sardinian inter-municipal
commuting network (SMCN): the nodes (red points) correspond to the towns, while the links to a flow value larger than 50 commuters between two towns.}
  \label{fig:1}
\end{figure}

Sardinia is the second largest Mediterranean island with an area of
approximately $24,000$ square kilometers and $1,600,000$ inhabitants. 
At the date of 1991, the
island was partitioned in $375$ municipalities, the second simplest body
in the Italian public administration, each one of those generally
corresponding to a major urban centre (in Figure \ref{fig:1} we
report the geographical distribution of the municipalities). 
For the whole set of municipalities the Italian National Institute of
Statistics \cite{istat} has issued the origin-destination table (OD)
corresponding to the commuting traffic at the inter-city level. The
OD is constructed on the output of a survey about commuting behaviors
of Sardinian citizens. This survey refers to the daily movement from
the habitual residence (the origin) to the most frequent place of employment (the destination): 
the data comprise both the
transportation means used and the time usually spent for
displacement. Hence, OD data give access to the flows of people
regularly commuting among the Sardinian municipalities. In particular
we have considered the external flows $i\to j$ which measure the movements
from any municipality $i$ to the municipality $j$ and we will focus on the
flows of individuals (workers and students) commuting throughout the
set of Sardinian municipalities by all means of transportation. This
data source allows the construction of the SMCN in which each node corresponds to a given
municipality and the links represent the presence of a non-zero flow
of commuters among the corresponding municipalities.

We are able then to construct a symmetric weighted
adjacency matrix $W$ in which the elements $w_{ij}$ are computed as
the sum of the $i\to j$ and $j\to i$ flows between the corresponding
municipalities (per day). The elements $w_{ij}$ are null in the case
of municipalities $i$ and $j$ which do not exchange commuting traffic
and by definition the diagonal elements are set to zero . According to
the assumption of regular bi-directional movements along the links,
the weight matrix is symmetric and the network is described as an
undirected weighted graph. 
The weighted graph provides a richer description since it considers the topology along with the
quantitative information on the dynamics occurring in the whole
network. 

\subsection{ARC Network}
The Atlanta Regional Commission maintains a network model for land use purposes of the metropolitan area of the city of Atlanta, in the State of Georgia, USA. 
The ARC travel demand model is designed to represent the state of the practice in travel demand modeling and to meet all modeling requirements in the US EPA Transportation Conformity Rule.
Further details on the arrangement of zones are reported in \cite{arc}.

The main data source for the calibration of the travel demand models was a household travel survey of eight thousand households conducted for the ARC from April 2001 through April 2002. 
The household survey data was the main source of data for developing the trip generation and distribution model. The trip generation model is a fairly unique trip based model in that it estimated the frequency a person will make trips, by the purpose of the trip, and then applies this frequency to individual persons to determine the total amount of travel made by the residents of the region.
Therefore, just like in the case of the SMCN network, the trips reported in the ARC model are produced by a trip generation model, which is calibrated according to the result of a survey. 
Further details are available in \cite{arc}.
The calibration is achieved by matching the trip length, frequency and by evaluating geographic area biases ({\it e.g.}, natural features, political or service delivery boundaries, etc).

The work presented in this paper is centered on the activity of commuters, which in the ARC model are described as Home Based Work (HBW) trips. 
It is commonplace to describe such trips as trips made for the purpose of work and which either begin or end at the traveler's home. 
This is a typical trip purpose that is related to the employment at the destination zone and population/household income of the traveler or the household at the origin zone.
Mode details on the nature and calibration of the HBW demand and distribution model can be found in \cite{arc} for this specific model. 
The nature of the relationship between demand for travel and land-use are further explored in the modeling review works by Wilson \cite{wilson} and Batty \cite{batty}.

\begin{figure}[h!]
\centering
  \includegraphics[width=9.0cm]{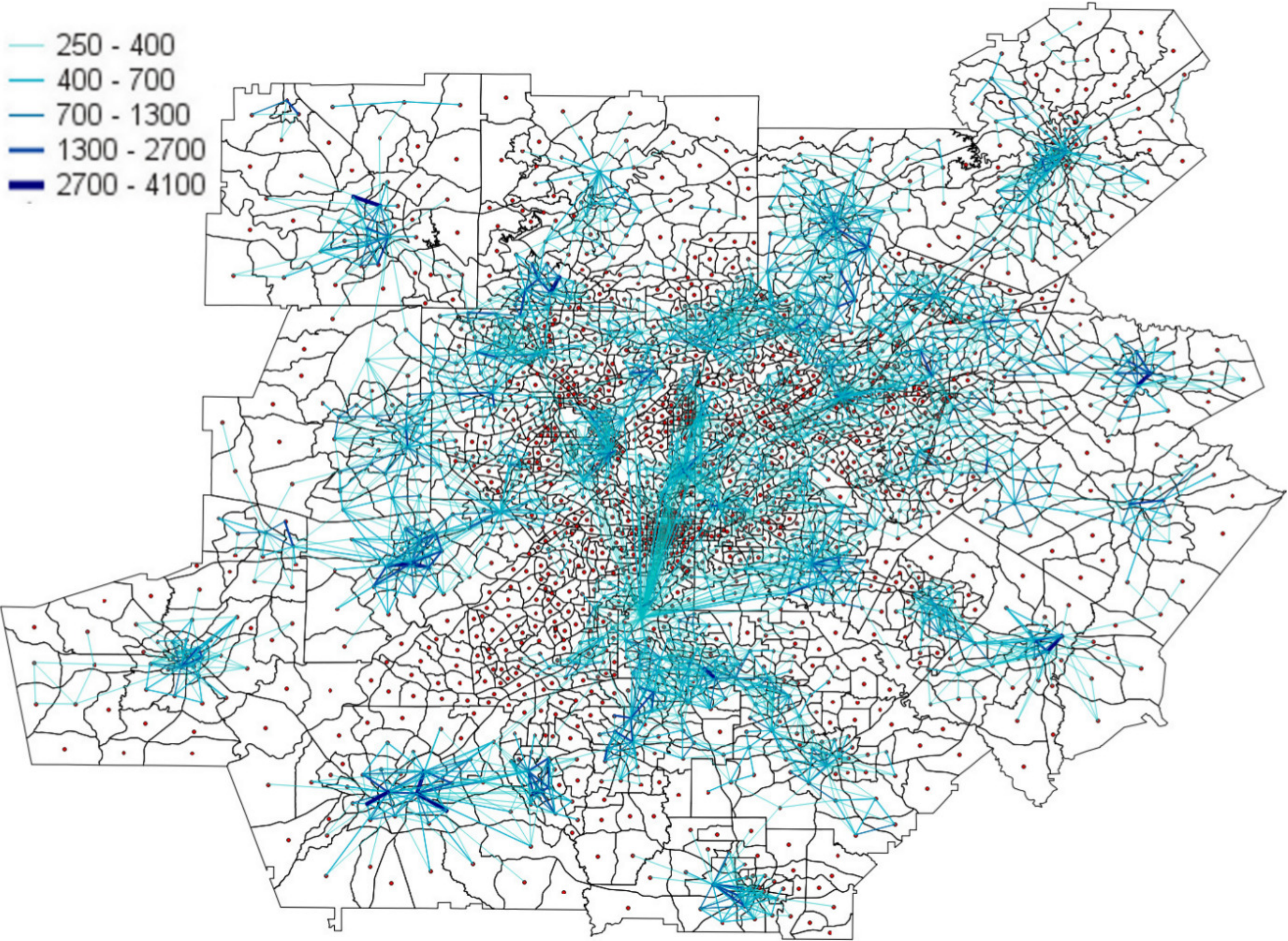}
  \caption{Extension of the zone system in the ARC model. Only the links with a weight greater than 250 have been shown. Each point is a centroid of a TAZ.}
  \label{fig:Partition}
\end{figure}

A number of socioeconomic variables are recorded in the ARC model, which are of importance for planning purpose and as inputs to the trip generation and demand growth algorithms. The figures below show, in order, the gradient plots of population and employment per zone, as recorded in the nationwide Census 2010. 
Darker zones indicate higher value for the corresponding variable.

\begin{figure}[h!]
\centering
  \includegraphics[width=9.0cm]{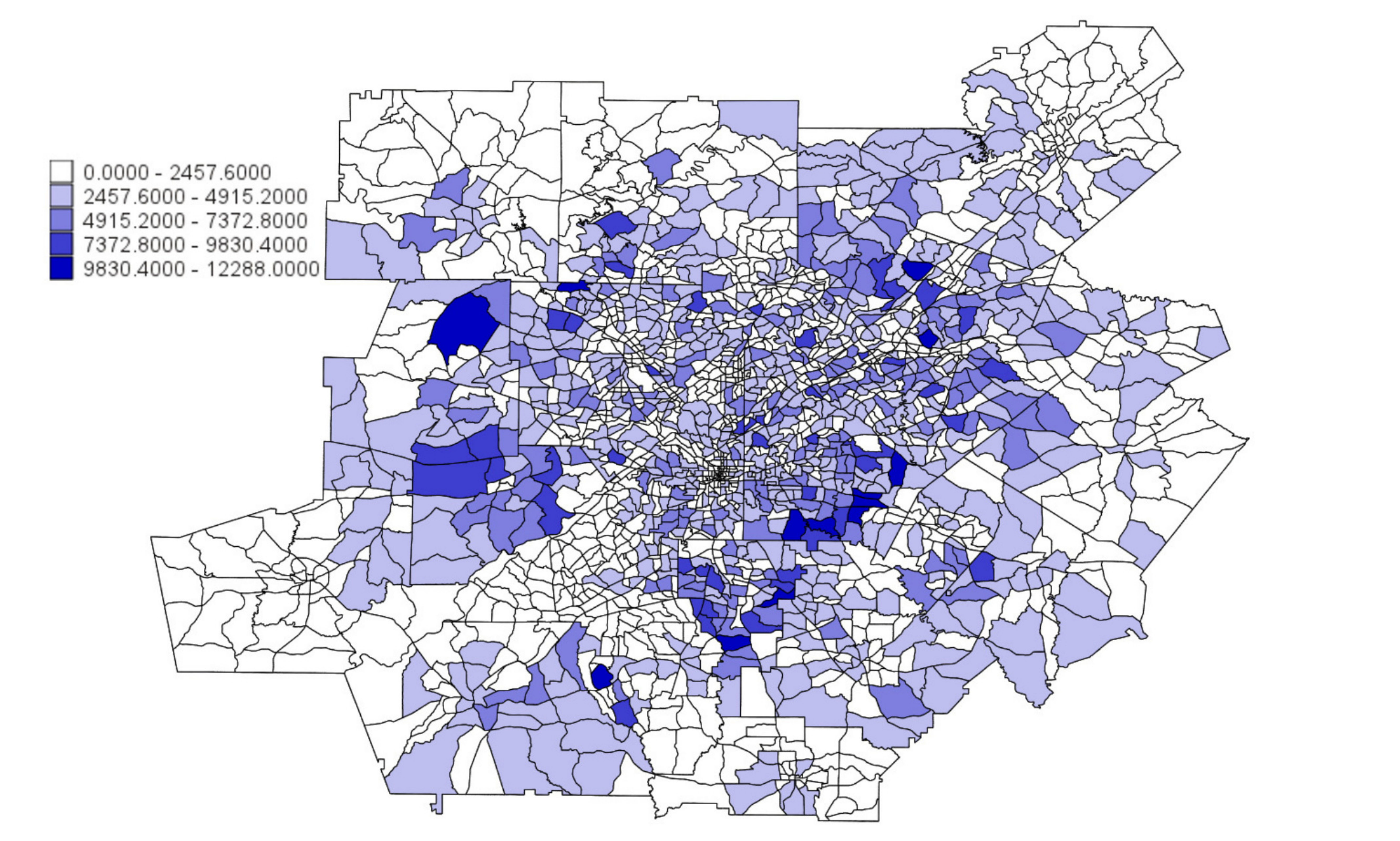}
  \caption{Gradient plot for Population in the ARC model.}
  \label{fig:popolazione}
\end{figure}
\begin{figure}[h!]
\centering
  \includegraphics[width=9.0cm]{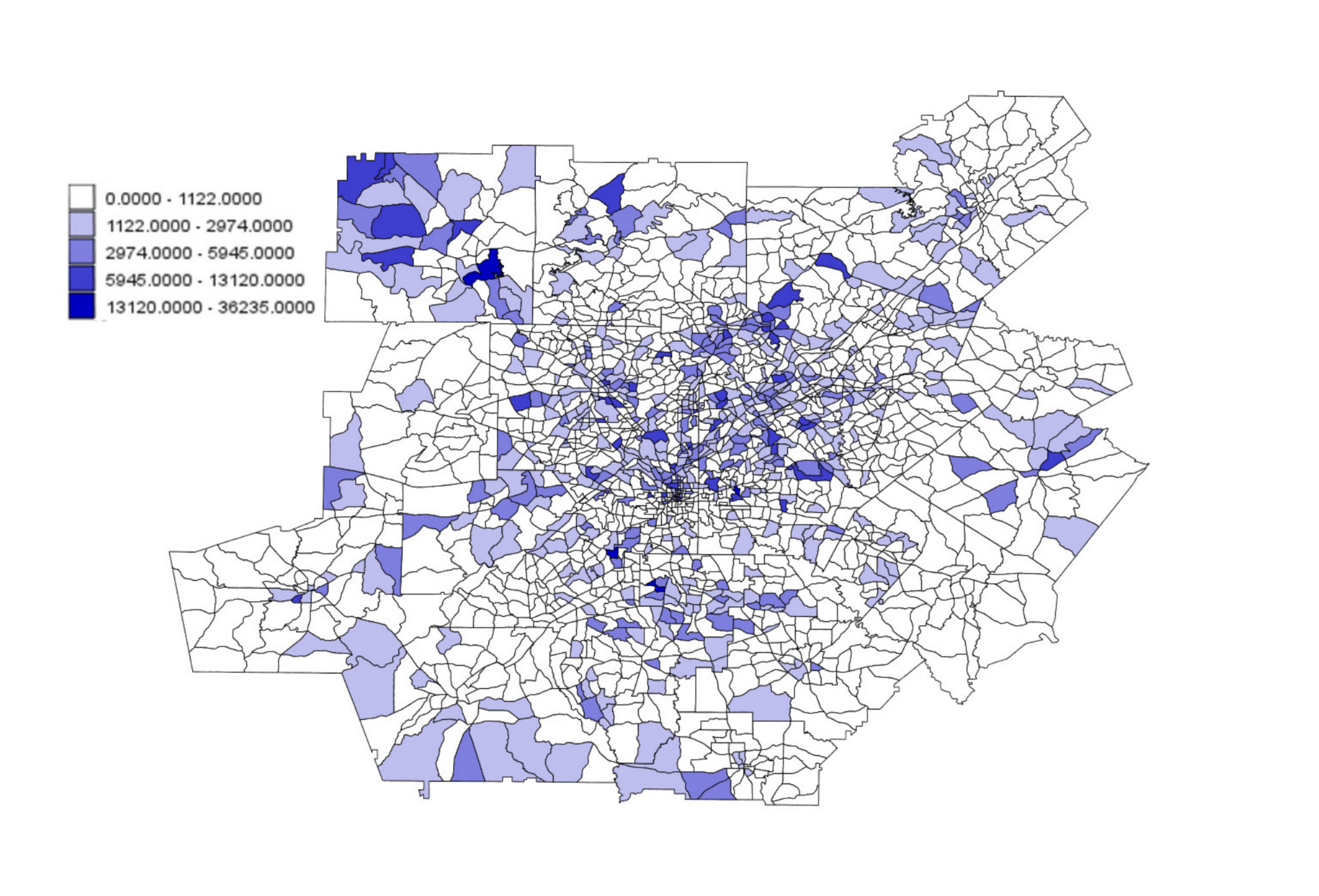}
  \caption{Gradient plot for Employment in the ARC model.}
  \label{fig:lavoro}
\end{figure}
Figure \ref{fig:popolazione} shows the gradient plot of the zone population. Population is seen in this figure as being scattered around the center that forms the core of the downtown area. 

Figure \ref{fig:lavoro} shows the gradient plot for the zone employment, measured as the number of jobs located in the zone the variable refers to. 
Employment is seen in this figure as primarily located in the downtown zones (which are quite small in size) plus other job centers in the suburban metropolitan areas.

\section{Results}
The sequence of charts that follow describes the correlation of the quantity $dQ$ and the various socioeconomic variables that are available for analysis.

\begin{figure}[h!]
\centering
  \includegraphics[width=9.0cm]{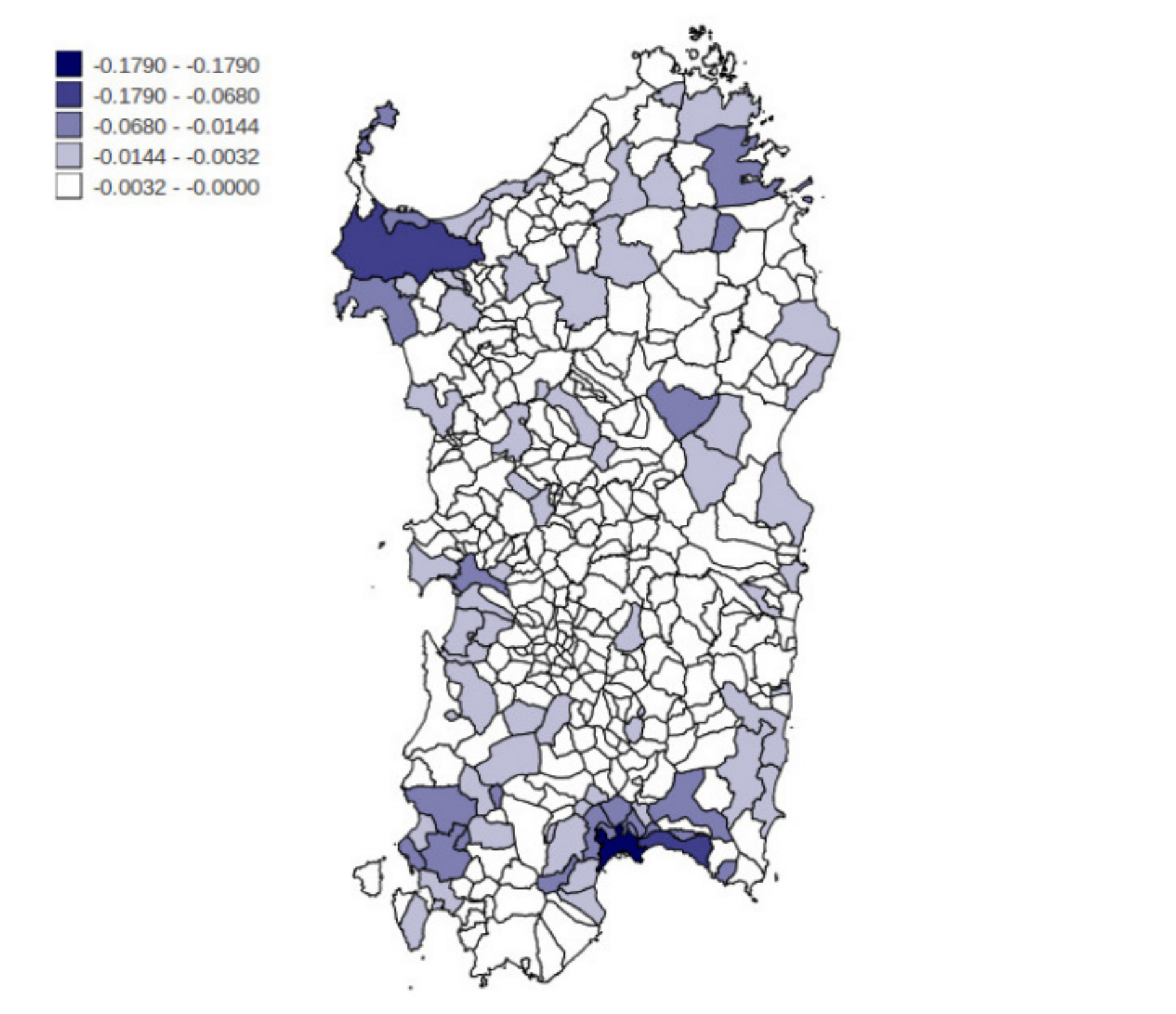}
  \caption{$dQ$ plot for the network related to Employment in the SMCN network.}
  \label{fig:sardegna_all_work_pendolari_dQ}
\end{figure}

\begin{figure}[h!]
\centering
  \includegraphics[width=9.0cm]{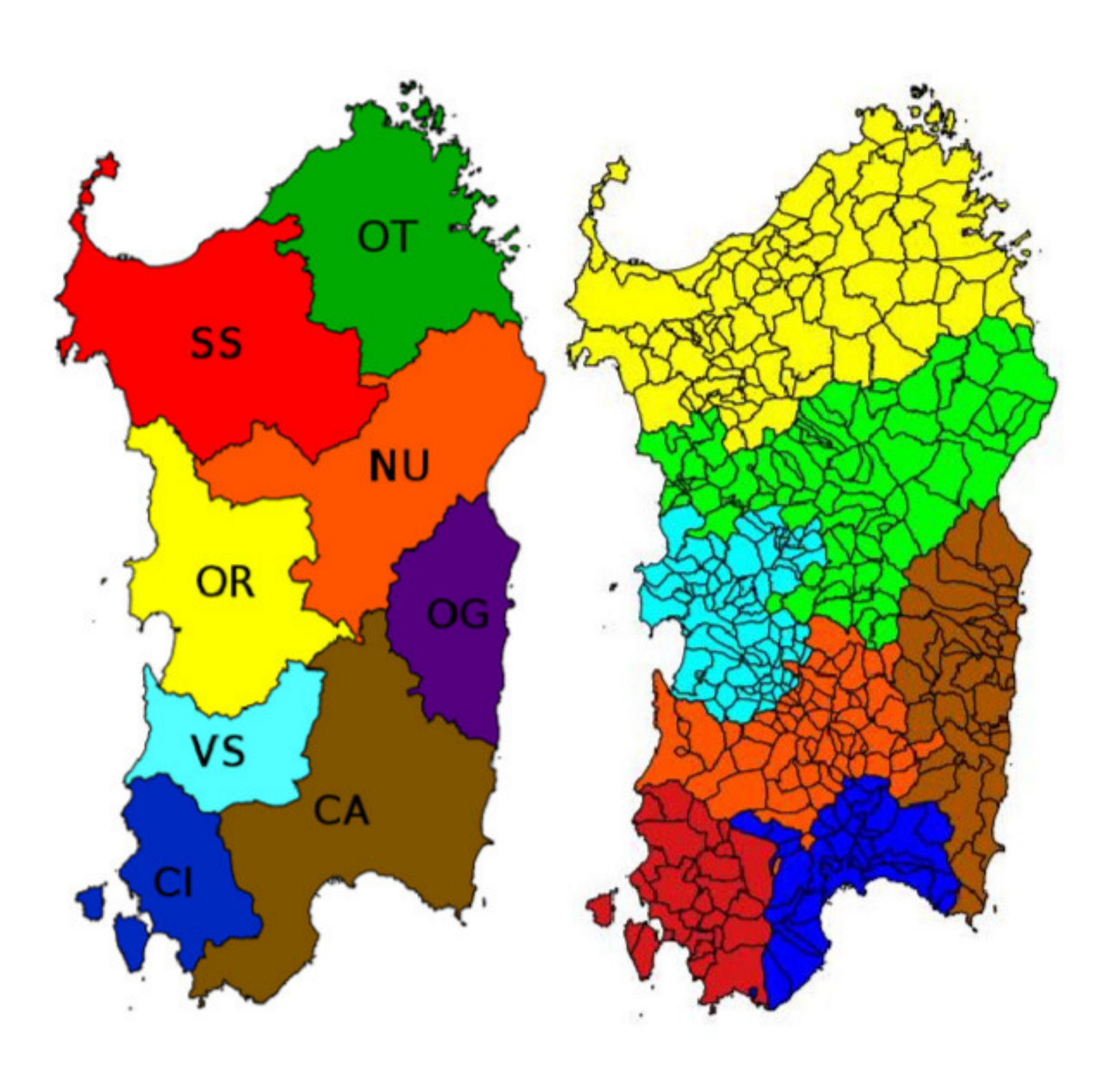}
  \caption{A comparison between the current provincial division (CA = Cagliari, CI = Carbonia-Iglesias, VS = Medio Campidano, OR = Oristano, OG = Ogliastra, 
    NU = Nuoro, SS = Sassari and OT = Olbia-Tempio) of the Sardinia region, Italy, and the result of the community detection.}
  \label{fig:sardegna_provincie_e_CD_topologica}
\end{figure}

The table below shows the result of correlation analysis between the computed $dQ$ and the in-strength of the various zones in the SMCN network. For the sake of clarity, the Sardinian and ARC networks are directed, as previously described in \ref{sec:Datasets}, and the in-strength has been computed starting from these original networks. However, the community detection has been performed using undirected networks obtained from the directed ones by summing up the weigths of incoming and outgoing links. 
\begin{table}[t]
\begin{tabular}{|l|c|c|}
\hline
Network & in-strength & Employment \\
\hline \hline
SMCN & 0.984 & 0.984 \\ 
ARC & 0.782 & 0.520 \\ 
\hline 
\end{tabular} 
\caption{Results of correlation analysis between $dQ$ and the in-strength and Employment.}
\label{ALL-correlation}
\end{table}
The correlation results shown in the table \ref{ALL-correlation} only give a overall picture of the quality of correlation between traffic and community structure. 
Figures \ref{fig:sardegna_all_work_pendolari_dQ}-\ref{fig:sardegna_provincie_e_CD_topologica} show the geographic distribution of the gradients of $dQ$ values across the zone system.
Figure \ref{fig:sardegna_all_work_pendolari_dQ} shows the values of $dQ$ arranged by color (darker color indicates higher value).
Higher $dQ$ indicates that the zone under investigation is more to the center of a community than the zones with lighter color. 
The data in Figure \ref{fig:sardegna_all_work_pendolari_dQ} shows that the two likeliest centers of a community (the two darkest zones in the figure) are not both centers of population and/or employment, nor are all large centers of population and/or employment necessarily key zones to the definition (and for its definition, stability) of a community.
In other words, community and socioeconomic activity are not on a one-to-one relationship, and it is not always possible to imply a ranking of one of these quantities with respect to the other and viceversa.

Figure \ref{fig:sardegna_provincie_e_CD_topologica} (right) below shows what the communities identified look like with respect to the political subdivisions of the island of Sardinia, the provinces that corresponds to the NUT3 regions in the international classifications (left). 
To put this result in context, it is important to note that the present political subdivision in eight provinces took effect in 2005 after a law passed in 2001 raised the number of provinces from the original number of four. 
Therefore, at the time the ISTAT data was collected (2001), Sardinia was subdivided politically in four provinces, hence the results of the modularity analysis showed that at least seven communities existed, subdivided geographically roughly along the lines of the boundary of the new (and present time) provinces.
The two subdivisions, "topological" the first, political the second, are remarkably alike, suggesting that either the political subdivision was designed to accomodate the arrangement of commuting movements, or the topological subdivision is a result of ease of movement within a (not yet established) political subdivision.

Finally, it is worth noting that, according to the results of a regional referendum in May 2012, the four new provinces established in according to the 2001 law will be abolished starting March 2013.

\begin{figure}[h!]
\centering
  \includegraphics[width=10.0cm]{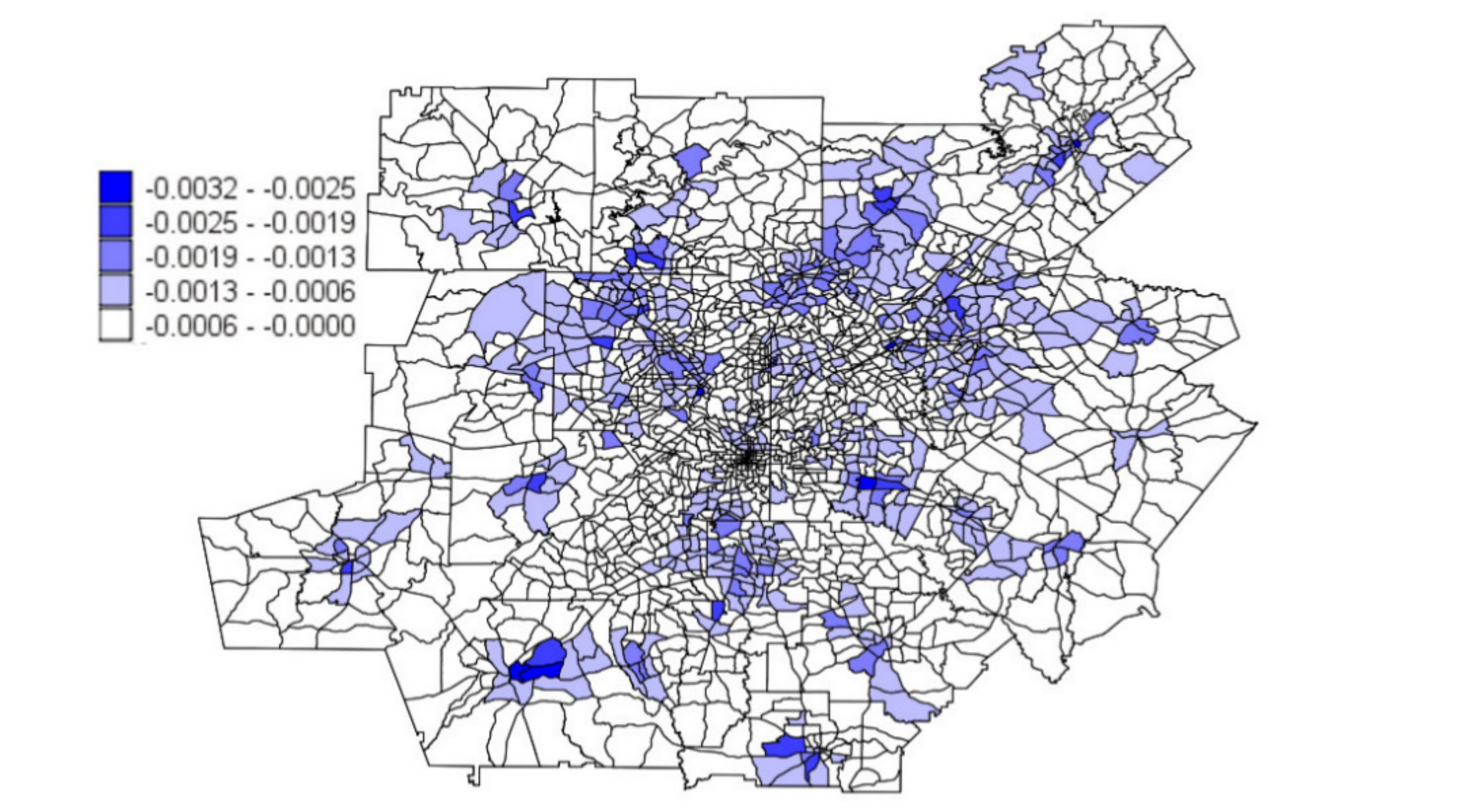}
  \caption{$dQ$ plot for the ARC network.}
  \label{fig:dQ}
\end{figure}

\begin{figure}[h!]
\centering
  \includegraphics[width=10.0cm]{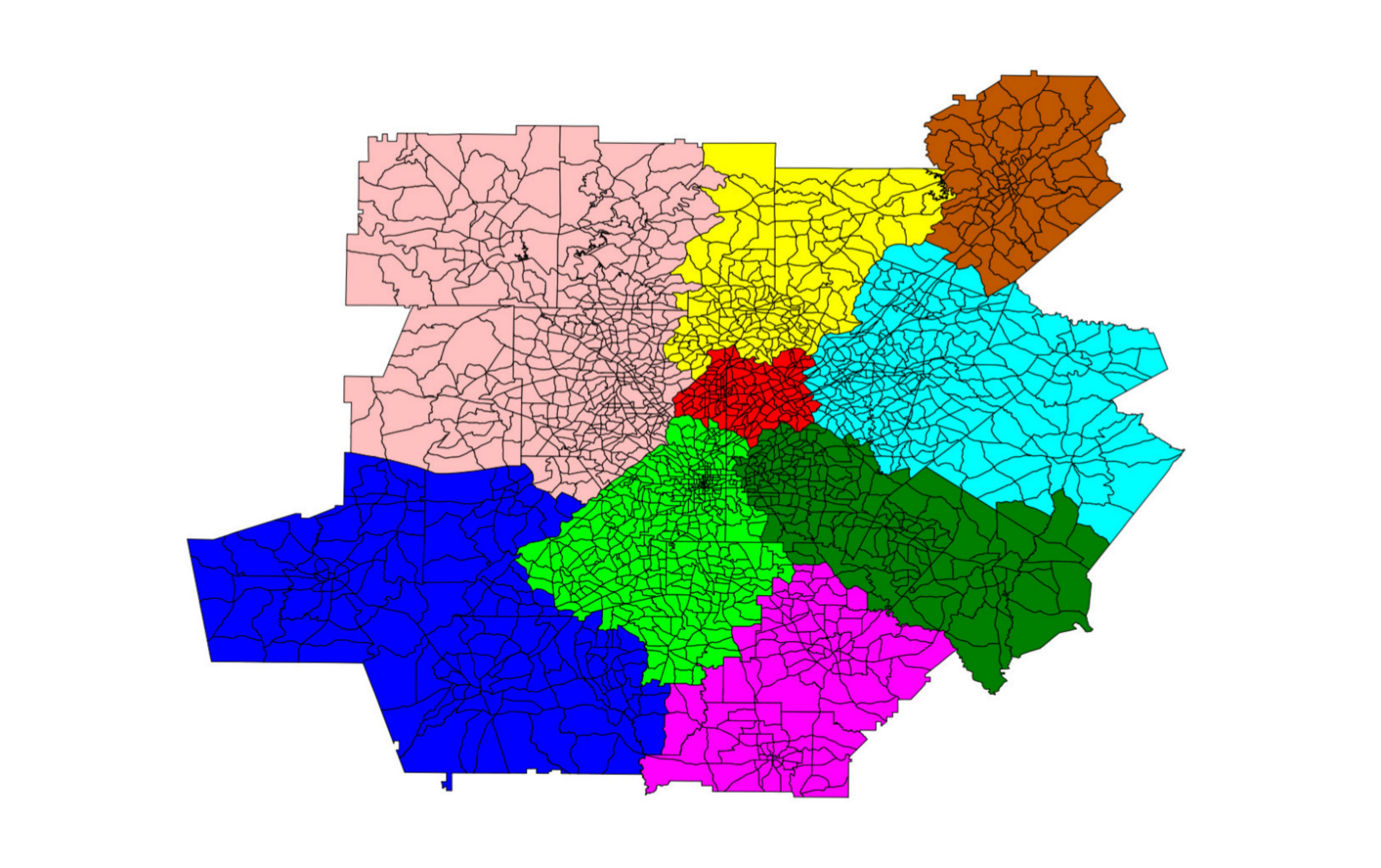}
  \caption{$dQ$ and community boundary plot for the ARC network}
  \label{fig:dQ_and_partition}
\end{figure}

Table \ref{ALL-correlation} shows also the result of the correlation between in-strength, $dQ$ and employment for the ARC network. 
Correlation with employment is poorer, while as in the case of the SMCN network, correlation with the in-strength is quite good. 
It is instructive then to see the geographic arrangement of the communities and other features of the network.
Figure \ref{fig:dQ} shows the $dQ$ distribution for the ARC network.
Darker zones indicate zones with higher $dQ$, and the darkest zones can be considered as the center of a community.
Figure \ref{fig:dQ_and_partition} show (color-coded) the community boudaries. 
\begin{figure}[h!]
\centering
  \includegraphics[width=9.0cm]{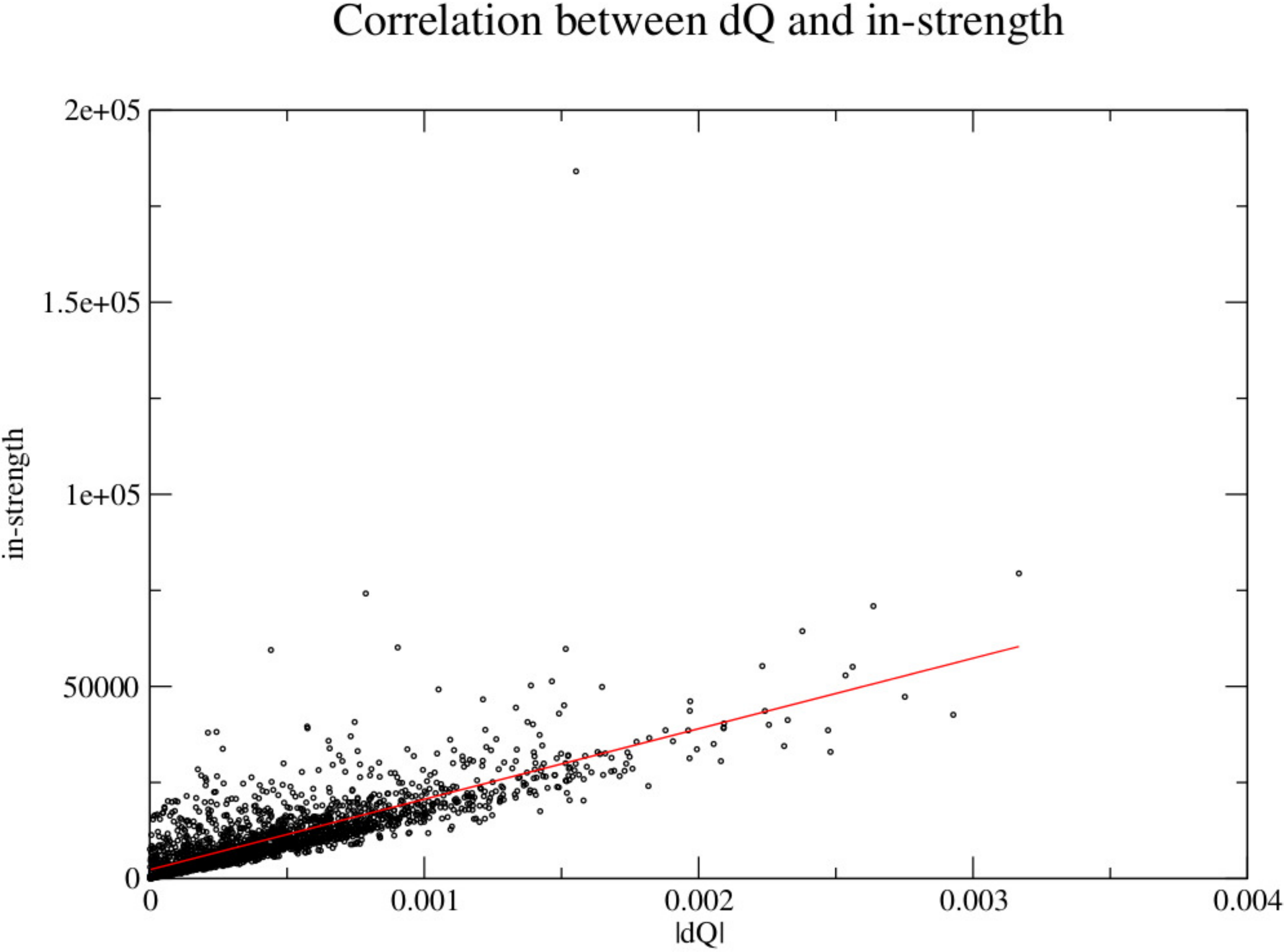}
  \caption{The correlation between dQ and in-strength is equal to 0.78.}
  \label{fig:dQ_vs_instrength}
\end{figure}
The correlation between $dQ$ and in-strength is explored by means of the Figure \ref{fig:dQ_vs_instrength}, which shows a correlation of almost 0.8.

\section{Discussion}
As per the case of the networks considered, community and socioeconomic structure are not on a one-to-one relationship, and it is not always possible to imply a ranking of one of these quantities with respect to the other and viceversa.
This conclusion is validated by visual comparison of Figures \ref{fig:dQ} and \ref{fig:lavoro}. 
While it appears reasonable that the communities be defined by the density of employment in a geographic area, the comparison between the two figures show that the community centers (defined by the highest $dQ$) are not necessarily arranged around the employment centers. 
A better fit is shown by comparing \ref{fig:dQ} with \ref{fig:Partition}, which is reflected in the better correlation between $dQ$ and in-strength, however even here it is possible to note smaller discrepancies.

Additionally, it appears from the correlation analysis previously described that in the SMCN network, the correlation with both in-strength and employment is much higher than in the case if the ARC network. 
While correlation with in-strength is good in both networks, correlation with employment is lower in the ARC network. This lower correlation may be explained in part with the size of the zones in the two networks. 
As we mentioned earlier, the zone system in the SMCN network is at the municipality level, while in the ARC network the size of the zones changes according to the structure of the road network, and they are generally smaller in size, down to the size of a traffic block, in the downtown areas. 
Therefore, some home-based-work (HBW) trips and viceversa may be "composite" trips, {\it e.g.}, a home-to-work trip that is in fact composed of a home-to-else trip plus a else-to-work trip in the same timeframe. 
As this disaggregation takes place more likely within the same municipality, this disaggregation is more likely to be present in a trip dataset based on a very granular zone system.

The two case studies that have been the subject of this analysis showed that community structure coming from the network analysis with its cores definitions, and socioeconomic structure are not on a one-to-one relationship, and it is not always possible to imply a ranking of one of these quantities with respect to the other and viceversa.
Hence, the "community" is a distinct mathematical object with its own land-use meaning that contains some valuable information not yet exploited. 
A “community”  in this sense is a subset of zones of the whole zone system that are linked together by having a frequency of HBW movements most likely where origin and destination are within the same subset or community. From a planning perspective, network modifications on a specific node within a community will have primarily large effects over its own community, of magnitude related to the $dQ$ of that zone. Effects on communities that do not contain that node are minimal.
It appears evident then why such description leaves room for modification of the boundary of a community. This aspect is in line with the previous comment that a community is defined by the algorithm that determines it.
Correlation between the community stability (expressed in $dQ$ value) and socioeconomic variables only tells part of story, while the remaining contribution to the community stability is to be found in the topological property of the networks. Our application to transportation networks has been a kind of territorial benchmark for this novel approach, but the proposed method for detecting cores in communities through the optimization of the modularity function is quite general and can be applied to other networked systems.



\begin{acknowledgments}
The authors wish to thank The Atlanta Regional Commission for the availability of the regional travel demand model. AC acknowledges support from FET Open project
255987 FOC and CNR PNR project “CRISIS Lab”.
FC gratefully acknowledges Sardinia Regional Government for the financial support of her PhD scholarship (P.O.R. Sardegna F.S.E. Operational Programme 
of the Autonomous Region of Sardinia, European Social Fund 2007-2013 - Axis IV Human Resources, Objective l.3, Line of Activity l.3.1.)
Finally, GS wishes to thank Prof. Franco Meloni and the Dipartimento di Fisica of the University of Cagliari for the kind hospitality.

\end{acknowledgments}





\begin{thebibliography}{300}
\bibitem{Fortunato:2010} S. Fortunato, "Community detection in graphs," {\em Physics Reports}, {\bf 486}, 75--174 (2010).
\bibitem{fhwa} {\em Traffic Detector Handbook}, FHWA operations material, available at {\tt http://www.fhwa.dot.gov/}.
\bibitem{mishan} E.J. Mishan, {\em Cost-Benefit Analysis}, Praeger Publishers (1976).
\bibitem{pashigian} B.P. Pashigian, {\em Price Theory and Applications}, McGraw-Hill (1995).
\bibitem{flood} J. Sohn, "Evaluating the significance of highway network links under the flood damage: An accessibility approach," {\em Transportation Research A}, {\bf 40}, 419--506 (2006).
\bibitem{Albert_et:2000} R. Albert, H. Jeong and A.-L. Barab\'asi,"Error and attack tolerance of complex networks," {\em Nature}, {\bf 406}, 378--382 (2000).
\bibitem{Sole_et:2008} R.V. Sol\'e, M. Rosas-Casals, B. Corominas-Murtra and S. Valverde, "Robustness of the European power grids under intentional attack," {\em Phys. Rev. E}, {\bf 77}, 1--7 (2008).
\bibitem{Schneider_et:2011} C.M. Schneider, A. Moreira, J.S. Andrade, S. Havlin, and H.J. Herrmann, "Mitigation of malicious attacks on networks," {\em Proc. Nat. Acad. Sc.}, {\bf 108}, 3838--3841 (2011).
\bibitem{Newman:2004} M.E.J. Newman and M. Girvan, "Finding and evaluating community structure in networks," {\em Phys. Rev. E}, {\bf 69}, 026113 (2004).
\bibitem{Fortunato:2007} S. Fortunato and M. Barthelemy, "Resolution limit in community detection," {\em Proc Natl Acad Sci, USA}, {\bf 104}, 36--41 (2007).
\bibitem{Newman:2006} M.E.J. Newman, "Modularity and community structure in networks," {\em Proc Natl Acad Sci, USA}, {\bf 103}, 8577--8582 (2006).
\bibitem{Blondel:2008} V.D. Blondel, J.-L. Guillaume, R. Lambiotte and E. Lefebvre, "Fast unfolding of communities in large networks," {\em J. Stat. Mech.}, {\bf 10}, P10018 (2008).
\bibitem{istat} {\em Censimento generale della popolazione e delle abitazioni - matrice origine destinazione degli spostamenti pendolari della Sardegna}, Italian National Institute of Statistics (ISTAT) (1991).
\bibitem{arc} {\em The Travel Forecasting Model Set For the Atlanta Region - 2008 Documentation}, by the Atlanta Regional Commission (2008).
\bibitem{wilson} A.G. Wilson, "Land-use/Transport Interaction Models," {\em Journal of Transport Economics \& Policy}, {\bf 32}, 3--26 (1997).
\bibitem{batty} M. Batty, {\em Urban Modelling: Algorithms, Calibrations, Predictions}, Cambridge University Press (1976).
\end{thebibliography}
\end{document}